\documentclass[pra,twocolumn,showpacs,nofootinbib,aps,10pt]{revtex4-1}
\usepackage{graphicx}
\usepackage{hyperref}
\usepackage[section]{placeins}
\usepackage{amsmath}
\usepackage{amsthm}
\usepackage{amssymb}
\usepackage{dsfont}
\usepackage{bbold}
\usepackage{comment}
\usepackage{color}
\usepackage[table]{xcolor}

\newtheorem*{theorem*}{Theorem}
\newtheorem*{lemma*}{Lemma}
\newcommand{\Tr}{\text{Tr}}

\newcommand{\bit}{\begin{itemize}}
\newcommand{\eit}{\end{itemize}\par\noindent}
\newcommand{\ben}{\begin{enumerate}}
\newcommand{\een}{\end{enumerate}\par\noindent}
\newcommand{\beq}{\begin{equation}}
\newcommand{\eeq}{\end{equation}\par\noindent}
\newcommand{\beqa}{\begin{eqnarray}}
\newcommand{\eeqa}{\end{eqnarray}\par\noindent}
\newcommand{\beqn}{\begin{eqnarray}}
\newcommand{\eeqn}{\end{eqnarray}\par\noindent}

\makeatletter
\def\hlinewd#1{%
  \noalign{\ifnum0=`}\fi\hrule \@height #1 \futurelet
   \reserved@a\@xhline}
\makeatother

\begin{document}

\title{From the Kochen-Specker theorem to noncontextuality inequalities \\without assuming determinism}
\author{Ravi Kunjwal}
\affiliation{Optics \& Quantum Information Group, The Institute of Mathematical Sciences, C.I.T Campus, Taramani, Chennai 600 113, India}
\author{Robert W. Spekkens}
\affiliation{Perimeter Institute for Theoretical Physics, 31 Caroline Street North, Waterloo, Ontario Canada N2L 2Y5}


\begin{abstract}
The Kochen-Specker theorem
demonstrates that it is not possible to reproduce the predictions of quantum
theory in terms of a hidden variable model where the hidden variables assign a
value to every projector 
deterministically and noncontextually. A noncontextual value-assignment to a
projector is one that does not depend on which other projectors---the
context---are measured together with it.
Using a generalization of the notion of noncontextuality that applies to both
measurements and preparations,
we propose a scheme for deriving inequalities that test whether a given set of
experimental statistics is consistent with a noncontextual model. Unlike
previous inequalities inspired by the 
Kochen-Specker theorem, we do not assume that the value-assignments are
deterministic and therefore in the face of a violation of our inequality, the
possibility of salvaging noncontextuality
by abandoning determinism is no longer an option. 
Our approach is operational in the sense that it does not presume quantum
theory: a violation of our inequality implies the impossibility of a
noncontextual model for {\em any} 
operational theory that can account for the experimental observations, including
any successor to quantum theory.
\end{abstract}
\maketitle

Although measurements in quantum theory 
cannot, in general, be 
implemented simultaneously,
one can 
still ask whether the outcomes of such
incompatible measurements might be simultaneously well-defined 
within some deeper theory. To formalize 
this deeper theory we use the framework
of {\em ontological models}~\cite{harriganspekkens} which generalizes the notion
of a hidden variable model.
Contrary to na\"{i}ve impressions, it is possible to find models of this sort
that reproduce quantum predictions.
Problems only arise if one makes additional assumptions about the model. 
The Kochen-Specker theorem~\cite{KochenSpecker} famously derives a contradiction
from an assumption 
we 
term {\em KS-noncontextuality}.
Consider a set of quantum measurements, each represented by an orthonormal
basis, such that some rays are common to more than one basis.  
It is assumed that every {\em ontic state}---a complete specification of the
properties of the system, including values of hidden variables---assigns a
definite value to each ray, 0 or 1, {\em regardless of the basis (i.e. context) 
in which the ray appears}. If a ray is assigned the value 1 (0) by an ontic
state $\lambda$, 
the measurement outcome associated with that ray is predicted to occur with
probability 1 (0) when any measurement 
including the ray is implemented on the system in ontic state $\lambda$. It
follows that for every basis, precisely one
ray must be assigned the value 1 and the others the value 0.

The assumption that the ontic state assigns a deterministic outcome to each
measurement is the greatest shortcoming of the Kochen-Specker theorem. Recall
that determinism is not an assumption 
of Bell's theorem~\cite{Belllocality,Bell}. This is evident from derivations of
the Clauser-Horne-Shimony-Holt inequality~\cite{chsh}. Even in Bell's original 1964
article~\cite{Belllocality}, where deterministic assignments play an important
role, determinism is not assumed but
rather {\em derived} from local causality and the fact that quantum theory
predicts perfect correlations if the same observable is measured on the two
parts of a maximally entangled state (an argument
from Einstein, Podolsky and Rosen \cite{epr} that Bell simply recycled
\cite{belldeterminism}).
It was shown in Ref.~\cite{Spe05} that one can make a similar argument about 
determinism in
noncontextual models:
rather than assuming it, one can derive it from a generalized notion of
noncontextuality 
and from two facts about quantum theory: (i) the outcome of a 
measurement of some observable is perfectly predictable whenever the preceding
preparation is of an eigenstate of that observable, and (ii) the
indistinguishability, relative to all quantum
measurements, of different convex decompositions of the completely mixed state
into pure states.

Hence, in any proof of the Kochen-Specker theorem one can replace the assumption of determinism with 
the generalized notion of noncontextuality and the quantum prediction of
perfect predictability. If perfect predictability is indeed observed, then
in the face of the resulting contradiction, one must give up on
noncontextuality. This contrasts with earlier proofs where 
one could always salvage the generalized notion of noncontextuality by abandoning determinism.

Of course, no real experiment ever yields {\em perfect} predictability, so this
manner of ruling out noncontextuality is not robust to experimental error. 
Following ideas introduced in recent work~\cite{exptlpaper}, we show how to contend with the lack of perfect predictability of measurements and derive an experimentally-robust noncontextuality inequality for any uncolourability proof of the Kochen-Specker theorem.

\begin{figure}
 \includegraphics[scale=0.3]{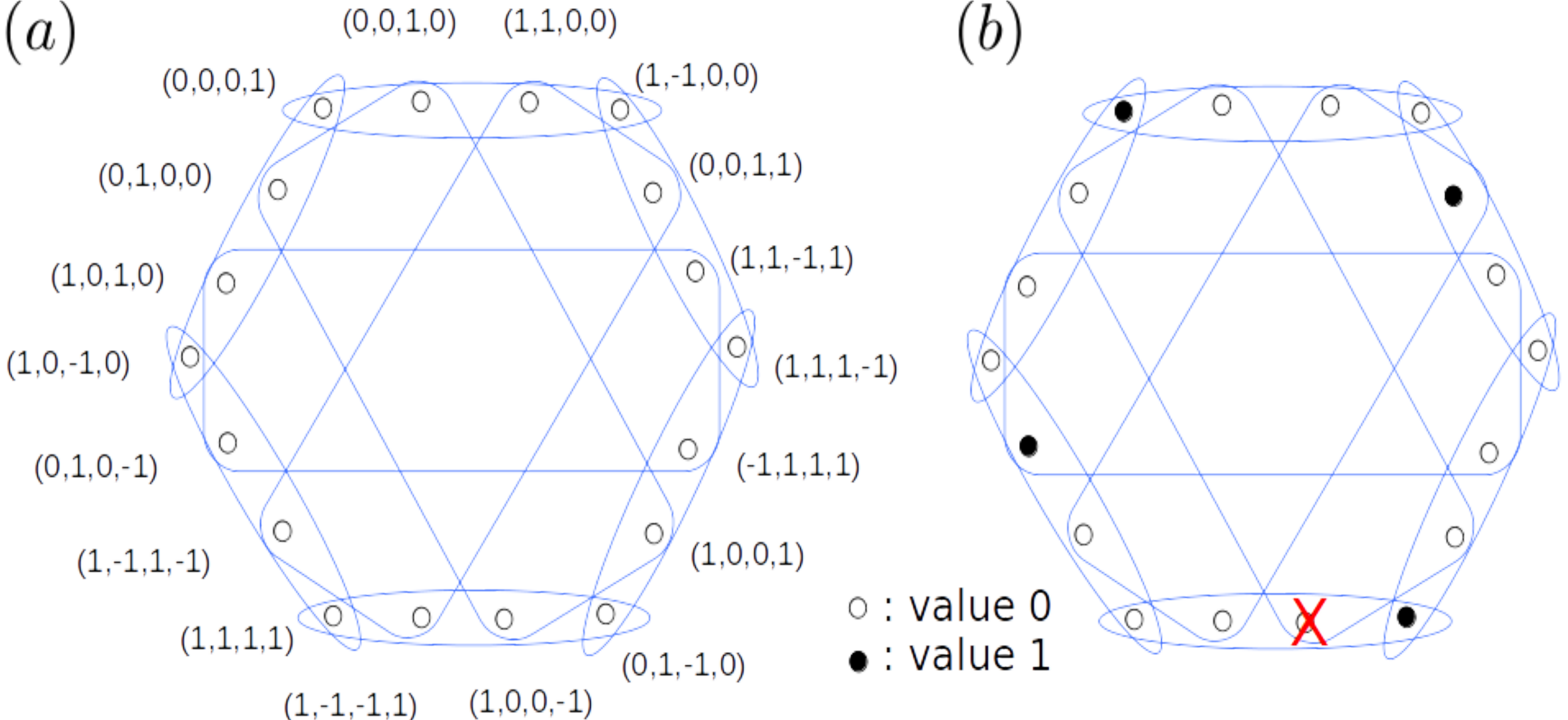}
 \caption{
Each of the 18 rays is depicted by a node, and the $9$ orthonormal bases 
are depicted by $9$ edges, each 
a loop encircling $4$ nodes. There is no noncontextual assignment of 0s and 1s to
these nodes 
such that for every edge precisely one node receives the value 1.
For instance, we have depicted a noncontextual assignment of 0s and 1s to 17 of
the rays, which cannot be completed to an assignment to all 18 rays because
neither 
value (0 or 1) can be assigned to the remaining ray (marked by X): while one basis in which it appears requires it to take the value 0,
the other requires the value 1.
}
\label{CEGAhypergraph}
\end{figure}

{\bf Review of the Kochen-Specker theorem.}  
The original proof of the KS theorem 
required 117 rays in a 3d Hilbert space~\cite{KochenSpecker}.
We use the much simpler proof in Ref.~\cite{Cabello18ray} as our illustrative
example.
It involves a 4d Hilbert space and 18 rays that appear in 9 orthonormal bases,
each ray appearing in two bases.  One can visualize this as a hypergraph 
with nodes 
representing the rays and 
edges 
representing orthonormal bases 
(Fig.~\ref{CEGAhypergraph}(a)). There is no
0-1 assignment to these rays that respects KS-noncontextuality: the
hypergraph is {\em uncolourable} (Fig.~\ref{CEGAhypergraph}(b)).
Of course, if the value assigned to a ray were allowed to be 0 in one basis and
1 in the other (a KS-contextual value assignment) then one could evade the
contradiction.

Is it possible to 
test the possibility of a
KS-noncontextual ontological model experimentally? One view is that the Kochen-Specker theorem
is not 
amenable to an experimental test. It merely constrains the
possibilities for {\em interpreting} the quantum formalism
\cite{MKC,merminquote}.
However, this answer is clearly inadequate. 
One {\em can} and {\em should} ask: what is the minimal set of operational
predictions of quantum theory that need to be experimentally verified in order
to show that it does not admit of a noncontextual model?

We show that this minimal set is a far cry from 
the whole of quantum theory and
is therefore consistent with many other possible operational theories. As such,
the no-go result we derive shows that none of these theories admit of a
noncontextual model.
Furthermore, if this set of predictions is corroborated by experiment, then this
implies that any future theory of physics that might replace quantum theory also
fails to admit of a noncontextual model.

We begin with some definitions.  An {\em operational theory} is a triple
$(\mathcal{P},\mathcal{M},p)$ where $\mathcal{P}$ is a set of preparations,
$\mathcal{M}$ is a set of measurements, and $p$ specifies, for every pair of
preparation and measurement, the probability distribution over outcomes for that
measurement if it is implemented on that preparation.  Specifically, if we
denote the set of outcomes of measurement $M$ by $\mathcal{K}_M$, then $\forall
P \in \mathcal{P},\; \forall M\in \mathcal{M}$, $p$ is a function of the form
 $p(\cdot|P,M):  \mathcal{K}_{M} \to [0,1]$.

An {\em ontological model} of an operational theory
$(\mathcal{P},\mathcal{M},p)$ is a triple $(\Lambda, \mu,\xi)$, where $\Lambda$
denotes a space of possible ontic states for the physical system (here presumed
to be discrete), where $\mu$ specifies a probability distribution over the ontic
states for every preparation procedure, that is, $\forall P\in \mathcal{P}, \;
\mu(\cdot|P): \Lambda \rightarrow [0,1]$, such that
$\sum_{\lambda\in\Lambda}\mu(\lambda|P)=1$, and where $\xi$ specifies, for every
measurement, the conditional probability of obtaining a given outcome if the
system is in a particular ontic state,  that is, $\forall M\in \mathcal{M},\;
\xi(k|M,\cdot):\Lambda \rightarrow [0,1]$, such that
$\sum_{k\in\mathcal{K}_M}\xi(k|M,\lambda)=1$.
In order for the ontological model to reproduce the statistical predictions of
the operational theory, it must be the case that
\beq
p(k|P,M)=\sum_{\lambda\in\Lambda} \xi(k|M,\lambda)\mu(\lambda|P)
\label{empirical}
\eeq
for all $P \in \mathcal{P}$, and $M\in \mathcal{M}$.

We denote the event of obtaining outcome $k$ of measurement $M$ by $[k|M]$.
If $[k|M]$
is assigned a deterministic outcome by every ontic state
in the ontological model, i.e., if $\xi(k|M,\cdot):\Lambda \rightarrow
\{0,1\}$,
then it is said to be {\em outcome-deterministic} in that model, and if this
holds for all $k$, then $M$ is also said to be outcome-deterministic.

We explain how to derive an experimental test of noncontextuality 
using a sequence of four refinements on the standard account of the KS theorem:

{\bf Operationalizing the notion of KS-noncontextuality.}
In a KS-noncontextual model of operational quantum theory, 
the value (0 or 1) 
assigned to the event $[k|M]$ by 
$\lambda$ is the same as the value 
assigned to the event $[k'|M']$
whenever these two events are represented by the same ray of Hilbert space
(here, we are assuming that $M$ and $M'$ are maximal projective measurements). 
We get to the crux of the notion of KS-noncontextuality, therefore, by
describing the {\em operational grounds} for associating the same ray to $[k|M]$
as is associated to $[k'|M']$.
 Letting $\Pi_{k|M}$ and $\Pi_{k'|M'}$ represent the corresponding rank-1
projectors, the grounds for concluding that $\Pi_{k|M}=\Pi_{k'|M'}$
 are that ${\rm tr}(\rho \Pi_{k|M})= {\rm tr}(\rho \Pi_{k'|M'})$ for an
appropriate set of density operators $\rho$.  It is clearly sufficient for the
equality to hold for the set of {\em all} density operators, but it is also
sufficient to have equality for certain smaller sets of density operators, namely, those
{\em complete for measurement tomography}, or simply {\em
tomographically complete}.

What then should the operational grounds be for assigning the same value to
$[k|M]$ and $[k'|M']$ in a general operational theory, where preparations are
not represented by density operators?  The answer, clearly, is that the event
$[k|M]$ {\em occurs with the same probability} as the event $[k'|M']$ for {\em
all} preparation procedures of the system,
\beq
p(k|M,P)=p(k'|M',P) {\rm\;\; for\; all\;\;  } P\in\mathcal{P},
\eeq
or equivalently, if this holds
for a subset of $\mathcal{P}$ that is tomographically complete.
In this case, we shall say that $[k|M]$ and $[k'|M']$ are {\em operationally
equivalent}, and denote this as $[k|M]$ $\simeq$ $[k'|M']$.  We can therefore
define a notion of KS-noncontextuality for any operational theory as follows: 
an ontological model $(\Lambda,\mu,\xi)$ of an operational theory
$(\mathcal{P},\mathcal{M},p)$ is KS-noncontextual if (i) operational equivalence
of events implies equivalent representations in the model, i.e., $[k|M]\simeq
[k'|M']\Rightarrow \xi(k|M,\lambda)=\xi(k'|M',\lambda)$  for all
$\lambda\in\Lambda$, and (ii) the model is outcome-deterministic,
$\xi(k|M,\cdot): \Lambda \to \{0,1\}.$

\begin{figure}[htb]
 \includegraphics[scale=0.3]{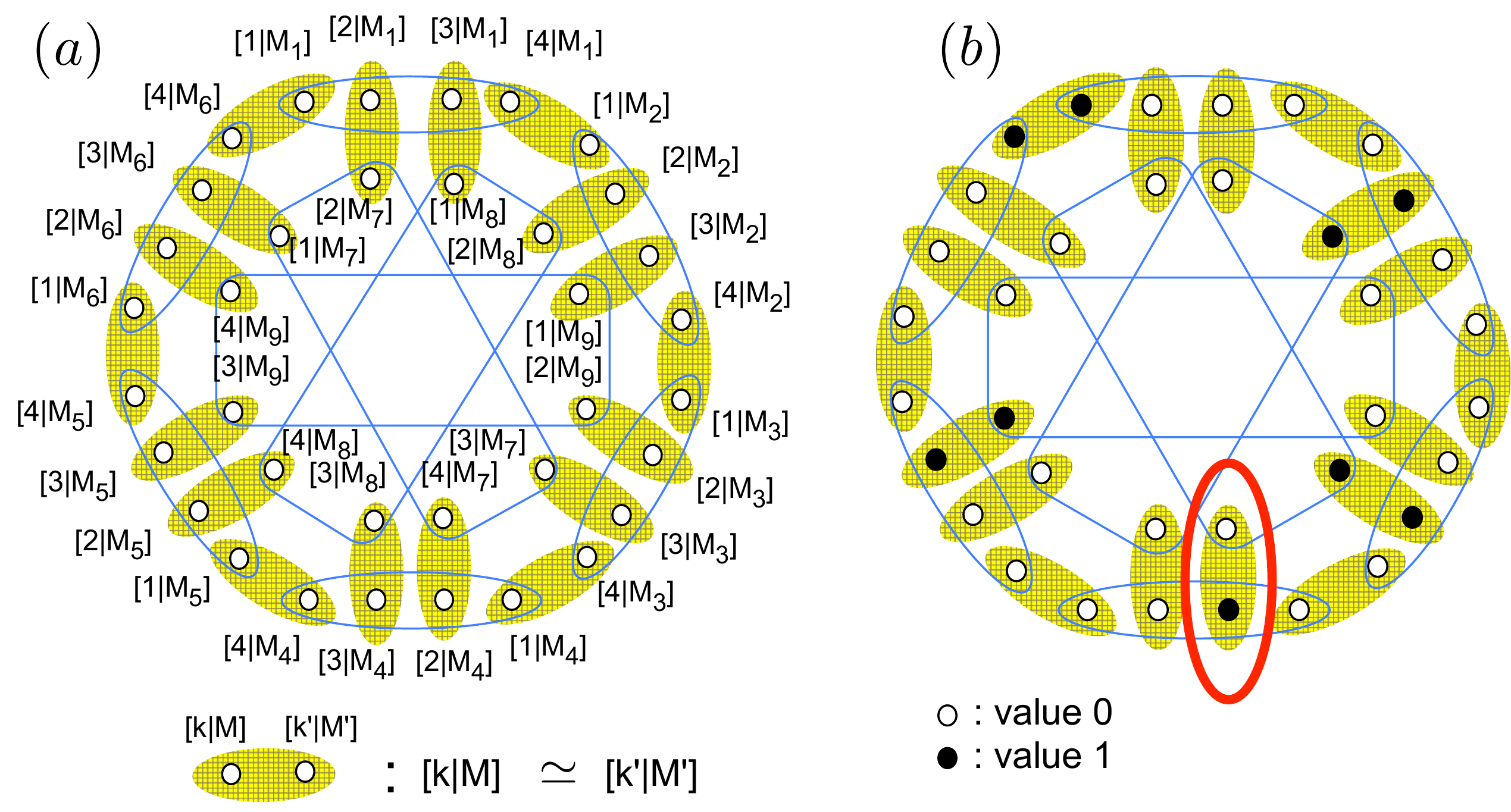}
 \caption{
(a) Nine four-outcome measurements. A blue loop encircling a set of
nodes implies that these nodes denote
outcomes of a single
measurement. A yellow hashed region enclosing a set of nodes 
implies that the corresponding events are operationally equivalent. (b) A
depiction of the fact that there is no outcome-deterministic noncontextual
assignment of values in $\{0,1\}$ to the measurements. The depicted value-assignment
breaks the assumption of noncontextuality for the pair of 
highlighted nodes
}
\label{mmtequivs}
\end{figure}

The operational equivalences among the measurements that are relevant for the 18
ray proof of the KS theorem depicted in Fig.~\ref{CEGAhypergraph}(a) are made
explicit in Fig.~\ref{mmtequivs}(a),
where 
every measurement event $[k|M]$ 
is represented 
by a distinct node, and a
novel type of edge between nodes 
specifies 
when two events are operationally
equivalent.
This representation affords a nice way of depicting contextual value assignments,
such as in Fig.~\ref{mmtequivs}(b).
It follows that 
{\em any} operational theory that
admits of nine four-outcome measurements that satisfy the operational
equivalence relations depicted in Fig.~\ref{mmtequivs}(a) fails to admit of a
KS-noncontextual model.

{\bf Defining a notion of noncontextuality without outcome determinism.}
The essence of 
noncontextuality is that context-independence at
the operational level should imply context-independence at the ontological
level. The operationalized version of KS-noncontextuality commits one to more than
this, however,
because it makes an additional assumption about {\em what sort of thing} should
be independent of context at the ontological level, namely, a deterministic
assignment of an outcome.
However, one can equally well assume
that the ontic state merely assigns a probability distribution over outcomes,
and take {\em this distribution} to be the thing 
independent of the
context.  
In Ref.~\cite{Spe05}, this 
revised notion of noncontextuality was
termed {\em measurement noncontextuality}:
\begin{quote}
\emph{Measurement noncontextuality} is satisfied by an ontological model
$(\Lambda,\mu,\xi)$ of an operational theory $(\mathcal{P},\mathcal{M},p)$ if
$[k|M]\simeq [k'|M']$ implies $\xi(k|M,\lambda)=\xi(k'|M',\lambda)$ for all
$\lambda\in \Lambda$.
\end{quote}
Here, 
$\xi(k|M,\cdot)\in[0,1]$
(and not merely $\{0,1\}$). Outcome determinism is not assumed.

{\bf Justifying outcome determinism for perfectly predictable measurements. }
Outcome determinism can, however, be justified sometimes
if one
assumes a notion of noncontextuality for {\em preparations}~\cite{Spe05}.
First, a definition: $P$ and $P'$ are said to be operationally equivalent,
denoted $P\simeq P'$, if for every measurement event $[k|M]$, $P$ assigns the
same probability to this event as $P'$ does, that is,
\beq
p(k|M,P)=p(k|M,P') {\rm\;\; for\; all\;\;  } k\in \mathcal{K}_M, {\rm\;\; for\;
all\;\;  } M\in\mathcal{M}.
\eeq
A preparation-noncontextual ontological model is then defined as
follows:
\begin{quote}
\emph{Preparation noncontextuality} is satisfied by an ontological model
$(\Lambda,\mu,\xi)$ of an operational theory $(\mathcal{P},\mathcal{M},p)$ if
$P\simeq P'$ implies $\mu(\lambda|P)=\mu(\lambda|P')$  for all
$\lambda\in\Lambda$.
\end{quote}
Insofar as both measurement and preparation noncontextuality are instances of
operational equivalence implying ontological equivalence, it is most natural to
assume {\em both}, that is, to assume {\em universal noncontextuality}.  

It was shown in 
Ref.~\cite{Spe05} that in a preparation-noncontextual model of quantum theory,
all projective measurements must be represented outcome-deterministically. 
Here, we provide a version of this argument for the 18 ray construction.

Suppose that one has experimentally identified thirty-six preparation procedures
organized into nine ensembles of four each, $\{ P_{i,k}: i \in \{ 1,\dots, 9\},
k\in \{1,\dots,4\}\}$, such that for all $i$, measurement $M_i$ on preparation
$P_{i,k}$ yields the $k$th outcome with certainty, 
\beq
\forall i , \forall k : p(k|M_i, P_{i,k})= 1.
\label{eq:perfectpredictability}
\eeq
We call this property {\em perfect correlation}.
In quantum theory, it suffices to let $P_{i,k}$ be the preparation associated
with the pure state corresponding to the $k$th element of the $i$th measurement
basis.

\begin{figure}
 \includegraphics[scale=0.3]{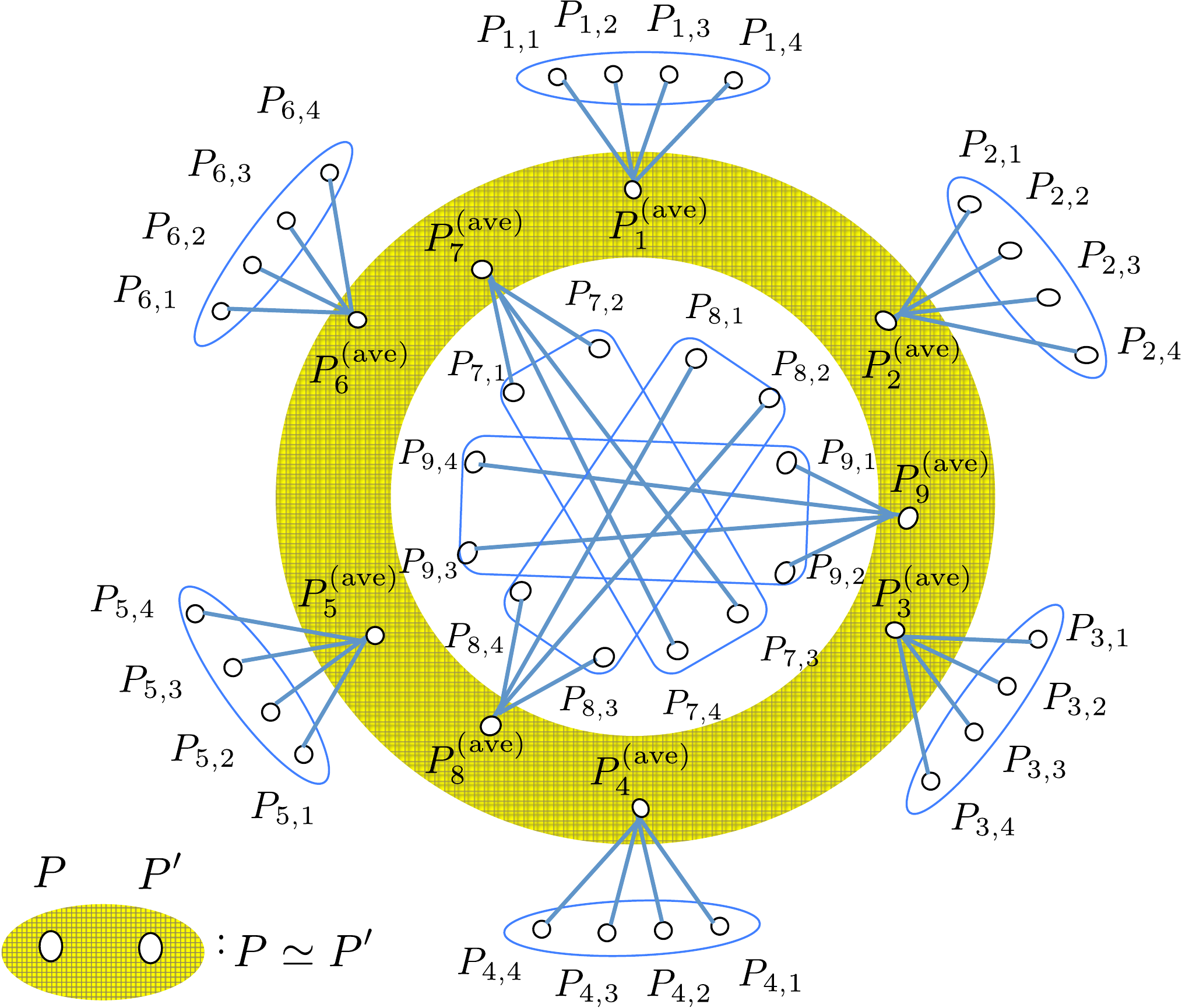}
 \caption{36 preparation procedures organized into nine ensembles of four each.
A node 
at the end of a set of lines emanating from the elements of an
ensemble represents 
the effective preparation procedure 
achieved by sampling uniformly from
the ensemble. A yellow region encircling a set of nodes implies that these
preparations are operationally equivalent.}
\label{prepequivs}
\end{figure}

Define the effective preparation $P_i^{\rm (ave)}$ as the procedure obtained by
sampling $k$ uniformly at random and then implementing $P_{i,k}$.
We now suppose that one has experimentally verified the operational equivalence
relations
\beq
P_i^{\rm (ave)} \simeq P_{i'}^{\rm (ave)} {\rm \;\; for\; all \;}
i,i'\in\{1,\dots,9\}.
\label{eq:prepequivs}
\eeq
These equivalences are depicted in Fig.~\ref{prepequivs}.  They hold in our
quantum example because the $P_i^{\rm (ave)}$ simply correspond to different
ways of preparing the completely mixed state. 

Given Eq.~\eqref{eq:prepequivs} and the assumption of preparation
noncontextuality, 
there is a single distribution over $\Lambda$,
denoted $\nu(\lambda)$, such that 
\beq
\mu(\lambda| P_i^{\rm (ave)})= \nu(\lambda)  {\rm \;\; for\; all \;} i
\in\{1,\dots,9\}.
\label{eq:nu}
\eeq
Given the definition of $P_i^{\rm (ave)}$, it follows that 
\beq
\frac{1}{4} \sum_k \mu(\lambda| P_{i,k})= \nu(\lambda)  {\rm \;\; for\; all \;}
i \in\{1,\dots,9\}.
\label{eq:nu2}
\eeq
Furthermore, recalling Eq.~\eqref{empirical}, for the ontological model to
reproduce Eq.~\eqref{eq:perfectpredictability}, we must have 
\beq
\forall i, \forall k: \sum_\lambda \xi(k|M_i,\lambda) \mu(\lambda| P_{i,k})= 1.
\label{eq:pp2}
\eeq
Because every $\lambda$ in the support of $\nu(\lambda)$ appears in the support
of $\mu(\lambda| P_{i,k})$ for some $k$, it follows that if $\xi(k|M_i,\lambda)$
had an indeterministic response
on any such $\lambda$, we would have a contradiction with Eq.~\eqref{eq:pp2}. 
Consequently, for all $i$ and $k$, 
the measurement event $[k|M_i]$ 
must be outcome-deterministic for all $\lambda$ in the support of
$\nu(\lambda)$.  

To summarize then, if one has experimentally verified the operational
equivalences depicted in Figs.~\ref{mmtequivs}(a) and \ref{prepequivs} and the
measurement statistics described in Eq.~\eqref{eq:perfectpredictability}, then
universal noncontextuality implies that the value assignments to measurement
events should be deterministic and noncontextual, hence KS-noncontextual, and we
obtain a contradiction in the usual manner. The argument can be
summarized thus
\beqa
&\textrm{universal  noncontextuality} + \textrm{operational
equivalences}\nonumber\\
&+ \textrm{perfect correlation}
\to \textrm{contradiction}.
\label{inference}
\eeqa

{\bf Contending with the lack of perfect predictability in real experiments.}
In real experiments, the ideal of perfect correlation described by
Eq.~\eqref{eq:perfectpredictability} is never achieved, so we cannot derive a
contradiction from it.  However, Eq.~\eqref{inference} is logically equivalent
to the following inference:
\beqa
&\textrm{universal  noncontextuality} + \textrm{operational
equivalences}\nonumber\\
&\to \textrm{failure of perfect correlation}.
\label{inference2}
\eeqa
This means that the amount of correlation, averaged over all $i$ and $k$, will
necessarily be bounded away from 1. 
It is this bound that is the operational noncontextuality inequality.
For the 18 ray example, we prove that
\begin{align}
A \equiv  \frac{1}{36} \sum_{i=1}^{9} \sum_{k=1}^{4} p(k|M_{i},P_{i,k}) \le
\frac{5}{6}.
\label{main}
\end{align}

To test the assumption of noncontextuality, therefore, one must measure the
correlation
$p(k|M_{i},P_{i,k})$ for all $i$ and $k$, but one must also verify that the
operational equivalences depicted in Figs.~\ref{mmtequivs}(a) and
\ref{prepequivs} hold, because only in this case 
does the assumption of noncontextuality imply that the inequality \eqref{main}
should hold.

We now outline how the bound in Eq.~\eqref{main} is obtained.  First, we use
Eq.~\eqref{empirical} to express $A$ in terms of $\xi(k|M_{i},\lambda)$ and
$\mu(\lambda|P_{i,k})$.  
Defining the {\em max-predictability} of a measurement $M$ given an ontic state
$\lambda$ by
 \beq
\zeta(M,\lambda)\equiv \max_{k'\in\mathcal{K}_M} \xi(k'|M,\lambda),
\label{eq:eta}
\eeq
we deduce that
\begin{eqnarray}
A &\le&\sum_{\lambda} \left( \frac{1}{9} \sum_{i}\zeta(M_{i},\lambda)
\left[ \frac{1}{4} \sum_{k}  \mu(\lambda|P_{i,k}) \right] \right)\nonumber \\
&=&  \sum_{\lambda} \left( \frac{1}{9} \sum_{i}\zeta(M_{i},\lambda) \right)
\nu(\lambda)\nonumber \\
&\le& \max_{\lambda} \left( \frac{1}{9} \sum_{i} \zeta(M_{i},\lambda)  \right),
\end{eqnarray}
where we have used Eq.~\eqref{eq:nu2}.

\begin{figure}
\includegraphics[scale=0.2]{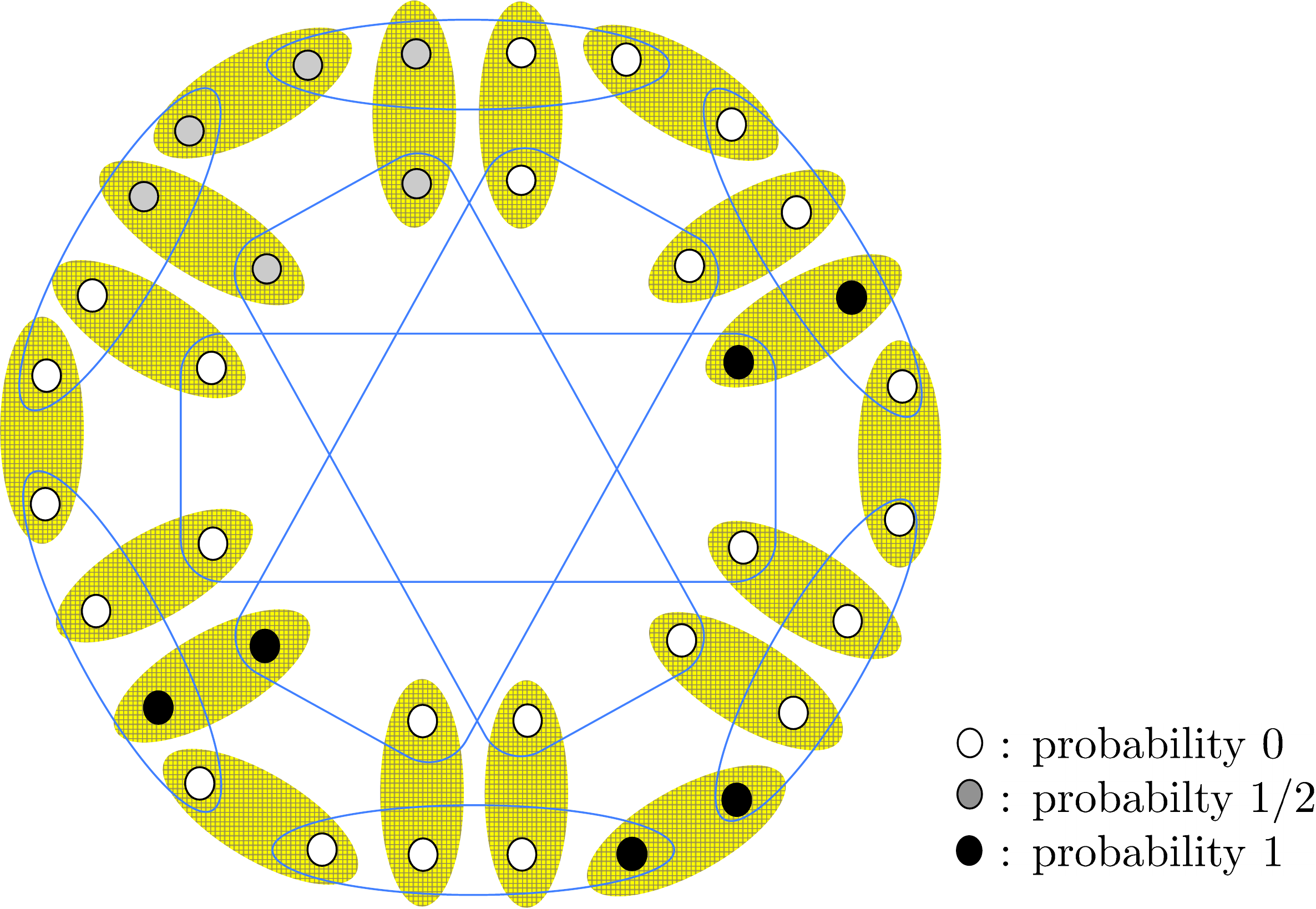}
\caption{Example of a noncontextual outcome-indeterministic assignment to the
measurements.
 }
\label{highpredictability}
\end{figure}

The measurements can
have indeterministic responses,
$\xi(k|M,\cdot):\Lambda \rightarrow [0,1]$, but measurement noncontextuality
implies that
$\xi(k|M_i,\lambda)=\xi(k'|M_{i'},\lambda)$ for the
operationally equivalent pairs $\{ [k|M_i], [k'|M_{i'}]\}$.
There are many such assignments.  Every unit-trace positive operator, for
instance, specifies an indeterministic noncontextual assignment via the Born
rule, and there are other, nonquantum assignments as well, such as the one
depicted in Fig.~\ref{highpredictability}.
Consider the average max-predictability achieved by the assignment of
Fig~\ref{highpredictability}.  Here, six measurements have max-predictability 1,
while three have max-predictability $\frac{1}{2}$. This implies that
$\frac{1}{9} \sum_{i} \zeta(M_{i},\lambda) = \frac{1}{9} ( 6 \cdot 1 + 3 \cdot
\frac{1}{2}) = \frac{5}{6}$.
As we demonstrate in Appendix \ref{proofineq}, no ontic state has a higher
average max-predictability than that of Fig.~\ref{highpredictability}, so that
$\max_{\lambda} \left( \frac{1}{9} \sum_{i} \zeta(M_{i},\lambda)  \right) \le
\frac{5}{6}$, thereby establishing the noncontextual bound on $A$.
The logical limit for the value of $A$ is $1$, so the noncontextual bound of
$\frac{5}{6}$ is nontrivial. The quantum realization of the 18 ray
construction
achieves $A=1$.

Note that if 
an experiment
fails to suppress noise sufficiently, then it may not succeed in
violating our noncontextuality inequality. This simple criterion of operational meaningfulness fails for
previous attempts at deriving noncontextuality inequalities \cite{Cabelloexpt},
a point we discuss further in Appendices \ref{noiserobust} and \ref{comparison}.
Although we have used the 18 ray
uncolourable set of Ref.~\cite{Cabello18ray} as an example, the scheme described can be
used to turn any proof of the Kochen-Specker theorem based on an uncolourable set into an experimental
inequality.
An issue we haven't addressed is that in practice no two measurement events are assigned {\em
exactly} the same probability by each of a tomographically complete set of preparations, nor do any two preparations assign
{\em exactly} the same probability distribution over outcomes to each of a tomographically complete set of measurements.
The solution to this problem is described in related work~\cite{exptlpaper, MattP}. 
A question that remains is: how does one accumulate evidence that 
a given set of measurements or preparations is indeed tomographically complete? 
This question represents the new frontier in the project of devising strict experimental tests of the assumption of noncontextuality.\color{black}

{\bf Acknowledgments}: RK thanks the Perimeter Institute and the Institute of
Mathematical Sciences for supporting his visit during the course of this work.
This project was made possible in part through the support of a grant from the
John Templeton Foundation.
Research at Perimeter Institute is supported by the Government of Canada through
Industry Canada and by the Province of Ontario through the Ministry of Economic 
Development and Innovation.

\appendix
\section{Proof of the inequality}\label{proofineq}

We can summarize our main result---a derivation of a noncontextuality inequality from the proof of the Kochen-Specker theorem for the 18 ray uncolourable set of Fig.~\ref{CEGAhypergraph}---by the following theorem:
\begin{theorem*}
Consider an operational theory $(\mathcal{P},\mathcal{M},p)$.  Let $\{ M_i \in \mathcal{M} :  i \in\{ 1,\dots,9\}\}$ be nine four-outcome measurements.
Let $[k|M_{i}]$ denote the $k$th outcome of the $i$th measurement, where $k \in \{1,\dots,4\}$. 
Let  $\{ P_{i,k} \in \mathcal{P}: i \in\{ 1,\dots,9\}$, $ k \in \{1,2,3,4\}\}$ be thirty-six preparation procedures, organized into nine sets of four.   Let $P^{\rm (ave)}_i\in \mathcal{P}$ be the preparation procedure obtained by sampling $k \in \{1,2,3,4\}$ uniformly at random and implementing $P_{i,k}$.

Suppose that one has experimentally verified the operational preparation equivalences depicted in Fig.~\ref{prepequivs}, namely,
 \begin{align}\label{eq:optlequivP1to9}
 P^{\rm (ave)}_1 \simeq P^{\rm (ave)}_2 \simeq \dots \simeq P^{\rm (ave)}_9,
\end{align} 
and the operational equivalences depicted in Fig.~\ref{mmtequivs}(a), namely,
 \begin{align}\label{eq:optlequivM18ray}
[k| M_{i}] \simeq [k'|M_{i'}],
\end{align} 
for the eighteen pairs specifed therein. 

If one assumes that the operational theory admits of a universally noncontextual ontological model,
that is, one which is both measurement-noncontextual and preparation-noncontextual, then the following inequality on operational probabilities holds
\begin{align}
A \equiv  \frac{1}{36} \sum_{i=1}^{9} \sum_{k=1}^{4} p(k|M_{i},P_{i,k}) \le \frac{5}{6}.
\end{align}
\end{theorem*}

We now provide the proof.  For clarity, we expand on some of the steps presented in the main article. 

Using Eq.~\eqref{empirical}, the quantity $A$ can be expressed in terms of the distributions and response functions of the ontological model as
\begin{align}
A = \frac{1}{36}\sum_{i=1}^{9}  \sum_{k=1}^{4} \sum_{\lambda}\xi(k|M_{i},\lambda)\mu(\lambda|P_{i,k}).
\end{align}
Using the definition of the max-probability $\zeta(M_{i},\lambda)$, given in Eq.~\eqref{eq:eta}, we have
\begin{align}
A \le \frac{1}{9}\sum_{i=1}^{9}  \sum_{\lambda}\zeta(M_{i},\lambda) \left( \frac{1}{4}\sum_{k=1}^{4} \mu(\lambda|P_{i,k}) \right).
\end{align}

Assuming that one experimentally verifies the operational preparation equivalences of Eq.~\eqref{eq:optlequivP1to9}, the assumption of preparation noncontextuality implies that
\beq
\mu(\lambda| P_1^{\rm (ave)})= \mu(\lambda| P_{2}^{\rm (ave)}) = \cdots = \mu(\lambda| P_{9}^{\rm (ave)}).
\eeq
It follows that there exists a single distribution, which we denote $\nu(\lambda)$, such that
\beq
\mu(\lambda| P_i^{\rm (ave)})= \nu(\lambda)  {\rm \;\; for\; all \;} i \in\{1,\dots,9\}.
\label{muPave}
\eeq
Recall that $P_i^{\rm (ave)}$ is the preparation procedure that samples $k$ uniformly from $\{1,2,3,4\}$ and implements $P_{i,k}$.  Given that the probability of the system being in a given ontic state $\lambda$ given the preparation $P_{i,k}$ is $\mu(\lambda|P_{i,k})$, and given that the probability of $P_{i,k}$ being implemented is $\frac{1}{4}$ for each value of $k$, it follows that the probability of the system being in a given ontic state $\lambda$ given the preparation $P_{i}^{\rm (ave)}$ is $\mu(\lambda|P_i^{\rm (ave)}) = \frac{1}{4} \sum_{\lambda} \mu(\lambda|P_{i,k})$.  Combining this with Eq.~\eqref{muPave}, we conclude that
\beq
\frac{1}{4} \sum_{\lambda} \mu(\lambda|P_{i,k}) = \nu(\lambda)  {\rm \;\; for\; all \;} i \in\{1,\dots,9\},
\eeq
and therefore that
\begin{align}\label{eq:aa}
A \le \frac{1}{9} \sum_{\lambda}\sum_{i=1}^{9} \zeta(M_{i},\lambda)\nu(\lambda).
\end{align}
This in turn implies
\begin{align}\label{eq:aa}
A \le \max_{\lambda}\frac{1}{9} \sum_{i=1}^{9} \zeta(M_{i},\lambda).
\end{align}

Assuming that one experimentally verifies the operational measurement equivalences of Eq.~\eqref{eq:optlequivM18ray}, the assumption of measurement noncontextuality implies that
\begin{align}
\xi(k|M_i,\lambda)=\xi(k'|M_{i'},\lambda),
\end{align}
for the eighteen pairs of operationally equivalent measurement events $([k|M_i], [k'|M_{i'}])$
specifed in Fig.~\ref{mmtequivs}(a). 

\begin{figure}[htb]
 \includegraphics[scale=0.30]{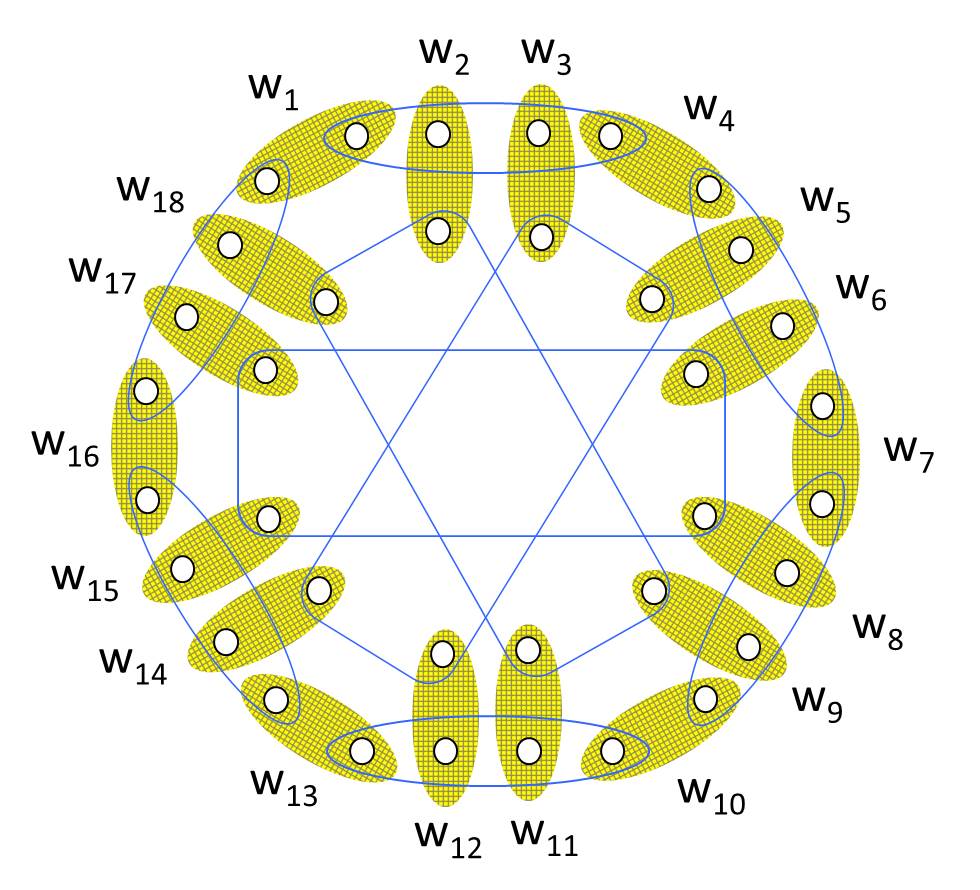}
 \caption{
 A choice of labelling of the eighteen equivalence classes of measurement events.  Here, $w_{\kappa}$ denotes the probability assigned to the equivalence class labelled by $\kappa$ in a noncontextual outcome-indeterministic ontological model.}
\label{legend}
\end{figure}

It is useful to simplify the notation at this stage.  We introduce the variable $\kappa \in \{1,\dots, 18\}$ to range over
the eighteen operational equivalence classes of measurement events.
We introduce the shorthand notation
\begin{equation}
w_{\kappa}\equiv \xi(k|M_i,\lambda)=\xi(k'|M_{i'},\lambda),
\end{equation}
for the probability assigned to the $\kappa$th equivalence class, where the dependence on $\lambda$ is left implicit.  The variable $\kappa$ enumerates the equivalence classes 
in Fig.~\ref{mmtequivs}(a) starting from 
$[1|M_1]$ and proceeding clockwise around the hypergraph, as depicted in Fig.~\ref{legend}. 

In this notation, the constraint that each response function is probability-valued, $\xi(k|M_i.\lambda) \in [0,1]$, is simply
\beq
0 \le w_{\kappa} \le 1, \;\;\forall \kappa \in\{1,\dots,18\},
\label{eq:aa}
\eeq
while the constraint that the set of response functions for each measurement sum to 1, $\sum_{k=1}^4 \xi(k|M_i,\lambda)=1$, can be captured by the matrix equality  
\beq\label{eq:bb}
Z \vec{w} =  \vec{u}
\eeq
where $\vec{w}\equiv (w_1,\dots,w_{18})^T$, $\vec{u}\equiv (1,1,1,1,1,1,1,1,1)^T$, and
\beq
Z \equiv \left(
\begin{array}{cccccccccccccccccc}
1&1&1&1&0&0&0&0&0&0&0&0&0&0&0&0&0&0\\
0&0&0&1&1&1&1&0&0&0&0&0&0&0&0&0&0&0\\
0&0&0&0&0&0&1&1&1&1&0&0&0&0&0&0&0&0\\
0&0&0&0&0&0&0&0&0&1&1&1&1&0&0&0&0&0\\
0&0&0&0&0&0&0&0&0&0&0&0&1&1&1&1&0&0\\
1&0&0&0&0&0&0&0&0&0&0&0&0&0&0&1&1&1\\
0&1&0&0&0&0&0&0&1&0&1&0&0&0&0&0&0&1\\
0&0&1&0&1&0&0&0&0&0&0&1&0&1&0&0&0&0\\
0&0&0&0&0&1&0&1&0&0&0&0&0&0&1&0&1&0
\end{array}
\right).
\eeq
Finally, we can express the quantity to be maximized as
\beq
\frac{1}{9} \sum_{i=1}^{9} \zeta(M_{i},\lambda) = \frac{1}{9} \sum_{i=1}^{9}  \max_{\kappa : Z_{i\kappa}=1} w_{\kappa},
\label{eq:cc}
\eeq
or, more explicitly, as
\begin{align}
&\frac{1}{9}\sum_{i=1}^{9} \zeta(M_{i},\lambda)\nonumber\\
&= \frac{1}{9} [ \max\{w_1,w_2,w_3,w_4\}
+ \max\{w_4,w_5,w_6,w_7\}\nonumber\\
&+\max\{w_7,w_8,w_9,w_{10}\}
+ \max\{w_{10},w_{11},w_{12},w_{13}\}\nonumber\\
&+ \max\{w_{13},w_{14},w_{15},w_{16}\}
+\max\{w_{16},w_{17},w_{18},w_1\}\nonumber\\
&+ \max\{w_{18},w_2,w_9,w_{11}\}
+ \max\{w_3,w_5,w_{12},w_{14}\}\nonumber\\
&+\max\{w_6,w_8,w_{15},w_{17}\} ].
\end{align}
The matrix equality of Eq.~\eqref{eq:bb} implies that there are only nine independent variables in the set $\{ w_1, w_2,\dots, w_{18} \}$ and that these satisfy linear inequalities.
The space of possibilities for the vector $\vec{w}$ therefore forms a nine-dimensional polytope in the hypercube described by Eq.~\eqref{eq:aa}.

The value of $\frac{1}{9} \sum_{i=1}^{9} \zeta(M_{i},\lambda)$ 
on any of the interior points of this polytope will be an average of its values at the vertices because it is a convex function of $\vec{w}$.  Therefore, to implement the maximization over $\lambda$, it suffices to maximize over the vertices of this polytope.

Following a brute-force enumeration of all the vertices of the polytope, the maximum possible value of $\frac{1}{9} \sum_{i=1}^{9} \zeta(M_{i},\lambda)$ is found to be $\frac{5}{6}$. 
An example of a vertex achieving this value is $\vec{w}=(1,0,0,0,1,0,0,0,\tfrac{1}{2},\tfrac{1}{2},\tfrac{1}{2},0,0,0,1,0,0,0)^{\rm T}$, which is depicted in Fig.~\ref{highpredictability}.  This concludes the proof.

Our proof technique can be adapted to derive a similar noncontextuality inequality correponding to any 
proof of the KS theorem based on the uncolourability of a set of rays of Hilbert space.   One begins by completing every set of orthogonal rays into a basis of the Hilbert space, and then forming the hypergraph depicting the orthogonality relations among these rays (the analogue of Fig.~\ref{CEGAhypergraph}).  One then forms the hypergraph decipting all of the measurements events, with one type of edge denoting which events correspond to the outcomes of a single measurement, and the other type of edge denoting when a set of measurement events are operationally equivalent (the analogue of Fig.~\ref{mmtequivs}(a)).  One then associates a set of preparations with every measurement in the hypergraph, one preparation for every outcome.  For each such set of preparations, we define the effective preparation that is the uniform mixture of the set's elements, and we presume that all of the effective preparations so defined are operationally equivalent (as is the case in quantum theory, where the effective 
preparation for every set corresponds to the completely mixed state).  We consider the correlation between the measurement outcome and 
the choice of preparation in the set associated with that measurement, averaged over all measurements.  This average correlation is the quantity $A$ that appears on the left-hand side of the operational inequality.  

The uncolourability of the hypergraph means that there are no noncontextual deterministic assignments
to the measurement events, hence the polytope of probabilistic assignments to the measurement events has no deterministic vertices either. Each vertex of this polytope, that is, each convexly-extremal probabilistic assignment, will necessarily yield an indeterministic assignment to some of the measurement events.  
Using the operational equivalences and the assumption of universal noncontextuality, one can infer from this 
that the average correlation $A$ is always bounded away from 1. 
For any uncolourable hypergraph, a quantum realization would achieve the logical limit $A=1$ by construction, so the noncontextuality inequality we derive is necessarily violated by quantum theory in each case.

One can understand this violation as being due to the fact that assignments of density operators that are independent of the  preparation context can achieve higher predictability for the respective measurements than assignments of probability distributions over ontic states that are independent of the preparation context. This is the feature of quantum theory that allows it to maximally violate the 
noncontextual bound of $A\leq 5/6$.

\section{Robustness of the noncontextuality inequality to noise}\label{noiserobust}
How much noise can one add to the measurements and preparations while still violating our noncontextuality inequality? We answer this question here assuming that the experimental operations are well-modelled by quantum theory.
According to quantum theory,
\begin{equation}
 p(k|M_i,P_{i,k})=\Tr(E_{k|M_i}\rho_{i,k}),
\end{equation}
where $E_{k|M_i}$ denotes the positive operator representing the measurement event $[k|M_i]$ and $\rho_{i,k}$  denotes the  density operator representing the preparation $P_{i,k}$.  To be precise, for every $i$, the set $\{ E_{k|M_i}\}_k$ is a positive operator valued measure, so that 
$0\leq E_{k|M_i}\leq I$, and $\sum_k E_{k|M_i}=I$, and for every $i$ and $k$, $\rho_{i,k}$ is positive, $\rho_{i,k}\geq 0$, and has unit trace, $\Tr\rho_{i,k}=1$.

In quantum theory, a noiseless and maximally informative measurement is represented by a POVM whose elements are rank-1 projectors, that is, 
\begin{equation}
 E_{k|M_i}=\Pi_{i,k},
\end{equation}
where for each $k$, $\Pi_{i,k}$ is a projector, hence idempotent, $\Pi_{i,k}^2=\Pi_{i,k}$, and is rank $1$, so that $\Pi_{i,k}=|\psi_{i,k}\rangle \langle \psi_{i,k}|$, where for each $i$, the set $\{ |\psi_{i,k}\rangle \}_k$ is an orthonormal basis of the Hilbert space.
If we furthermore set 
\beq
\rho_{i,k}=\Pi_{i,k},
\eeq 
then we find  $p(k|M_i,P_{i,k})={\rm Tr}(E_{k|M_i} \rho_{i,k})=1$ for each $(i,k)$, and consequently $A=1$.  We see, therefore, that the maximum possible value of $A$ is attained when measurements satisfy the noiseless ideal.  
We can now consider the consequence of adding noise. 

We begin by considering a very simple noise model wherein the preparations and measurements both deviate from the noiseless ideal by the action of a depolarizing channel, that is, a channel of the form
\begin{equation}
\mathcal{D}_p(\cdot) = p I (\cdot) I + (1-p) \frac{1}{4}I \;{\rm Tr}(\cdot),
\end{equation}
which with probability $p$ implements the identity channel and with probability $1-p$ generates the completely mixed state. 
If the quantum states are the image of the ideal states under a depolarizing channel with parameter $p_1$, and the POVM is obtained by acting the depolarizing channel with parameter $p_2$ followed by the ideal projector-valued measure (such that the POVM elements are the images of the projectors under the {\em adjoint} of the channel), then
\begin{eqnarray}
\rho_{i,k}&=&\mathcal{D}_{p_1} (\Pi_{i,k}) = p_1 \Pi_{i,k}+(1-p_1)\frac{1}{4}I,\\
E_{k|M_i}&=& \mathcal{D}^{\dag}_{p_2} (\Pi_{i,k})=  p_2 \Pi_{i,k}+(1-p_2) \frac{1}{4} I,
\end{eqnarray}
Here, the POVM $\{ E_{k|M_i}\}_k$ is a mixture of $\{ \Pi_{i,k} \}_k$ and a POVM $\{ \frac{1}{4}I,\frac{1}{4}I,\frac{1}{4}I,\frac{1}{4}I \}$ which simply samples $k$ uniformly at random regardless of the input state.
It follows that for each $(i,k)$, if we consider $p(k|M_i,P_{i,k})= {\rm Tr}(E_{k|M_i} \rho_{i,k})$, we find perfect predictability for the term having weight $p_1 p_2$ while for the three other terms, we have a uniformly random outcome, so that in all
\begin{equation}
 p(k|M_i,P_{i,k})=  p_1p_2 + (1-p_1p_2) \frac{1}{4}.
 \end{equation}
It follows that
\begin{equation}
A\equiv \frac{1}{36}\sum_{i=1}^9\sum_{k=1}^{4} p(k|M_i,P_{i,k})=\frac{1}{4}+\frac{3}{4}p_1 p_2,
\end{equation}
Thus a violation of the noncontextuality inequality, i.e. $A>\frac{5}{6}$, occurs if and only if
\begin{equation}
p_1 p_2 >\frac{7}{9}.
\end{equation}

It turns out that one can derive similar bounds for more general noise models as well.   Suppose that instead of a depolarizing channel, we have one of the form 
\begin{equation}
\mathcal{N}_{p,\rho}(\cdot) = p I (\cdot) I + (1-p) \rho \;{\rm Tr}(\cdot).
\end{equation}
With probability $p$, this implements the identity channel and with probability $1-p$ it reprepares a state $\rho$  that need not be the completely mixed state, but which is independent of the input to the channel.  The analogous sort of noise acting on the measurement corresponds to acting on the POVM elements by the adjoint of this channel, that is,
\begin{equation}
\mathcal{N}^{\dag}_{p,\rho}(\cdot) = p I (\cdot) I + (1-p) I \;{\rm Tr}(\rho \; \cdot).
\end{equation}

Therefore, if this sort of noise is applied to the ideal states and measurements, with the parameters in each noise model allowed to depend on $i$, we obtain
\begin{eqnarray}
\rho_{i,k}&=&\mathcal{N}_{p^{(i)}_1,\rho_i} (\Pi_{i,k}) = p^{(i)}_1 \Pi_{i,k}+(1-p^{(i)}_1) \rho^{(i)},\label{bb1}\\
E_{k|M_i}&=& \mathcal{N}^{\dag}_{p^{(i)}_2,\sigma_i} (\Pi_{i,k})=  p^{(i)}_2 \Pi_{i,k}+(1-p^{(i)}_2) s(k|i) I,\nonumber \\\label{bb2}
\end{eqnarray}
where $s(k|i)\equiv {\rm Tr}(\rho^{(i)} \Pi_{i,k})$ is a probability distribution over $k$ for each value of $i$.  
Here, the POVM $\{ E_{k|M_i}\}_k$ is a mixture of $\{ \Pi_{i,k} \}_k$ and a POVM $\{ s(k|i) I\}_k$ which simply samples $k$ at random from the distribution $s(k|i)$, regardless of the quantum state.  Compared to the simple model considered above, the innovation of this one is that for both preparations and measurements, the noise is allowed to be biased. 

For the case of $p_1^{(i)}=0$, which by Eq.~\eqref{bb1} implies that  $\rho_{i,k}=\rho^{(i)}$, we find that, regardless of the measurement, $p(k|M_i,P_{i,k})$ is just a normalized probability distribution over $k$ (because there is no $k$ dependence in the state).  Hence, in this case, $\frac{1}{4}\sum_{k=1}^4p(k|M_i,P_{i,k})= \frac{1}{4}$.

Similarly, for the case of $p_2^{(i)}=0$, that is, when the POVM corresponds to a random number generator $E_{k|M_i}=s(k|i) I$, we find that, regardless of the preparation, $p(k|M_i,P_{i,k})$ is again just a normalized probability distribution over $k$.  Hence, in this case again, $\frac{1}{4}\sum_{k=1}^4p(k|M_i,P_{i,k})= \frac{1}{4}$.  

It follows that for generic values of $p_1^{(i)}$ and $p_2^{(i)}$, we have $\frac{1}{4}\sum_{k=1}^4p(k|M_i,P_{i,k})= p_1^{(i)} p_2^{(i)} + (1-p_1^{(i)} p_2^{(i)}) \frac{1}{4}$.  In all then, we have 
\begin{equation}
A\equiv \frac{1}{36}\sum_{i=1}^9\sum_{k=1}^{4} p(k|M_i,P_{i,k})=\frac{1}{4}+\frac{3}{4} \left( \frac{1}{9}\sum_{i=1}^9 p^{(i)}_1 p^{(i)}_2 \right).
\end{equation}
Consequently, a violation of the noncontextuality inequality, i.e., $A>\frac{5}{6}$, occurs if and only if the noise parameters satisfy
\begin{equation}
\frac{1}{9}\sum_{i=1}^9 p^{(i)}_1 p^{(i)}_2  >\frac{7}{9}.
\end{equation}
Because the parameters $p_1^{(i)}$ and $p_2^{(i)}$ decrease as one increases the amount of noise, this inequality specifies an upper bound on the amount of noise that can be tolerated if one seeks to violate the noncontextuality inequality. 

This analysis highlights how the approach to deriving noncontextuality inequalities described in this article has no trouble accommodating noisy POVMs.  This contrasts with previous proposals for experimental tests based on the traditional notion of noncontextuality, which can only be applied to projective measurements.  This is one way to see how previous proposals are not applicable to realistic experiments, where every measurement has some noise and consequently is necessarily {\em not} represented projectively.

\section{Comparison to other noncontextuality inequalities}\label{comparison}

We have proposed a technique for deriving noncontextuality inequalities from proofs of the Kochen-Specker theorem.  It is useful to compare our approach with one that has previously been proposed by Cabello~\cite{Cabelloexpt}.
We do so by explicitly comparing the two proposals in the case of the 18 ray construction of Ref.~\cite{Cabello18ray}.   Indeed, the fact that Ref.~\cite{Cabelloexpt} proposes an inequality for this construction 
is part of our motivation for choosing it as our illustrative example.

For each of the eighteen operational equivalence classes of measurement events, labelled by $\kappa \in \{1,\dots,18\}$ as depicted in Fig.~\ref{legend}, we associate a $\{-1,+1\}$-valued variable, denoted $S_{\kappa} \in \{-1,+1\}$.
A given ontic state $\lambda$ is assumed to assign a value to each $S_{\kappa}$.
The fact that there is only a {\em single} variable associated to each equivalence class implies that any assignment of such values is necessarily noncontextual.

Ref.~\cite{Cabelloexpt} considers a particular linear combination of expectation values of products of these variables: 
\begin{align}
\alpha \equiv &-\langle S_1 S_2 S_3 S_4\rangle - \langle S_4 S_5 S_6 S_7\rangle- \langle S_7 S_8 S_9 S_{10}\rangle\nonumber\\
& - \langle S_{10} S_{11} S_{12} S_{13}\rangle -  \langle S_{13} S_{14} S_{15} S_{16}\rangle - \langle S_{16} S_{17} S_{18} S_{1}\rangle\nonumber\\
&- \langle S_{18} S_2 S_9 S_{11}\rangle- \langle S_3 S_5 S_{12} S_{14}\rangle\nonumber\\
&-\langle S_6 S_8 S_{15} S_{17}\rangle,
\end{align}
and derives the following inequality for it:
\begin{align}
\alpha \le 7
\label{Cabelloinequality}
\end{align}
(Note that Ref.~\cite{Cabelloexpt} used a labelling convention for the eighteen measurement events that is different from the one we use here; to translate between the two conventions, it suffices to compare Fig. 1 in that article with Fig.~\ref{legend} in ours.)  Each term in $\alpha$ refers to a quadruple of variables that can be measured together, that is, which can be computed from the outcome of a single measurement.   Different terms correspond to measurements that are incompatible.

In Ref.~\cite{Cabelloexpt}, the following justification is given for the inequality \eqref{Cabelloinequality}.  We are asked to consider 
the $2^{18}$ possible assignments to $(S_1,\dots, S_{18})$ that result from the two possible assignments to $S_{\kappa}$, namely $-1$ or $+1$, for each $\kappa \in \{1,\dots,18\}$.
It is then noted that among all such possibilities, the maximum value of $\alpha$ that can be achieved is 7.

Ref.~\cite{Cabelloexpt} states that a violation of this inequality should be considered evidence of a failure of noncontextuality.  We disagree with this conclusion, and the rest of this section seeks to explain why.

\subsection{The most natural interpretation}

\begin{figure}[htb]
 \includegraphics[scale=0.26]{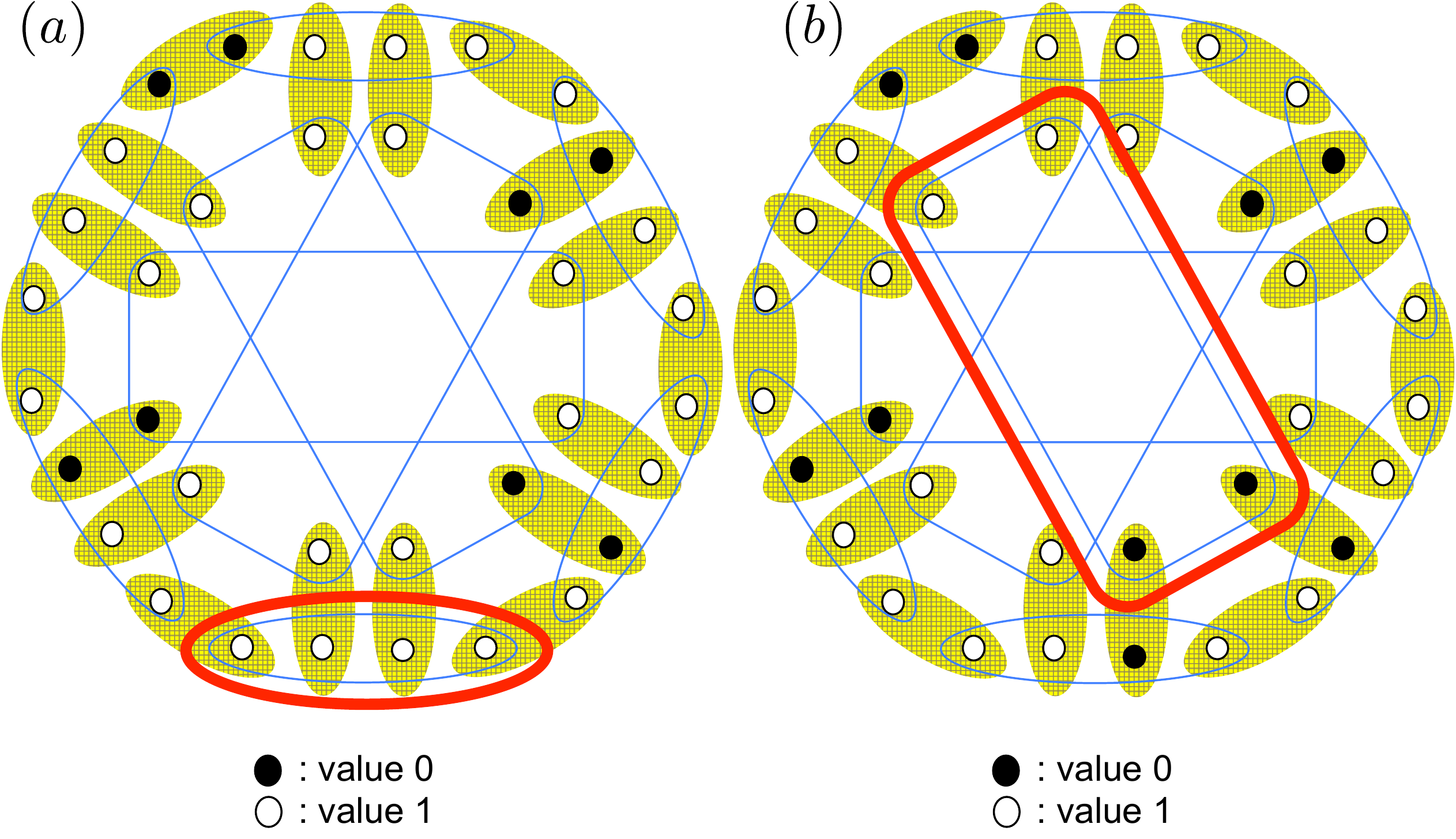}
 \caption{Examples of noncontextual assignments of $\{ 0,1\}$-values to the measurement events in Fig.~\ref{mmtequivs}(a) where it is not required that every measurement has precisely one outcome that is assigned value 1 and three outcomes that are assigned the value 0.  Example (a) depicts an assignment wherein there is a measurement all of whose outcomes receive probability 0.  Example (b) depicts one wherein there is a measurement two of whose outcomes recieve probability 1. }
\label{LogicDefyingAssignments}
\end{figure}

It is useful to recast the inequality of Eq.~\eqref{Cabelloinequality} in terms of variables $v_{\kappa}$ with values in $\{0,1\}$ rather than $\{-1,+1\}$.  Specifically, we take
\beq
v_{\kappa} \equiv \frac{S_{\kappa} +1}{2}.
\eeq
Under this translation, products of the $S_{\kappa}$ correspond to sums (modulo 2) of the $v_{\kappa}$.  For instance,  an equation such as $S_{\kappa_1} S_{\kappa_2}=-1$ corresponds to the equation $v_{\kappa_1} \oplus v_{\kappa_2} =1$, where $\oplus$ denotes sum modulo 2, while $S_{\kappa_1} S_{\kappa_2}= +1$ corresponds to $v_{\kappa_1} \oplus v_{\kappa_2} =0$, so that $v_{\kappa_1} \oplus v_{\kappa_2} =  \frac{-S_{\kappa_1} S_{\kappa_2} +1}{2}$.  In particular, we also have
\beq
v_{\kappa_1} \oplus v_{\kappa_2} \oplus v_{\kappa_3}\oplus v_{\kappa_4} =  \frac{-S_{\kappa_1} S_{\kappa_2} S_{\kappa_3}S_{\kappa_4} +1}{2}
\eeq
or equivalently,
\beq
- S_{\kappa_1} S_{\kappa_2} S_{\kappa_3}S_{\kappa_4}= 2(v_{\kappa_1} \oplus v_{\kappa_2} \oplus v_{\kappa_3}\oplus v_{\kappa_4}) -1,
\label{jjj}
\eeq 
We can therefore consider a quantity $\alpha'$, defined as
\begin{align}
\alpha' \equiv &\langle v_1 \oplus  v_2 \oplus v_3 \oplus v_4\rangle + \langle v_4 \oplus v_5 \oplus v_6 \oplus v_7\rangle\nonumber\\
&+ \langle v_7 \oplus v_8 \oplus v_9 \oplus v_{10}\rangle
 + \langle v_{10} \oplus v_{11} \oplus v_{12} \oplus v_{13}\rangle\nonumber\\
 & +  \langle v_{13} \oplus v_{14} \oplus v_{15} \oplus v_{16}\rangle +\langle v_{16} \oplus v_{17} \oplus v_{18} \oplus v_{1}\rangle\nonumber\\
&+ \langle v_{18} \oplus v_2 \oplus v_9 \oplus v_{11}\rangle + \langle v_3 \oplus v_5 \oplus v_{12} \oplus v_{14}\rangle\nonumber\\
&+ \langle v_6 \oplus v_8 \oplus v_{15} \oplus v_{17}\rangle,
\label{alphaprime}
\end{align}
so that $\alpha = 2\alpha' - 9$, 
and we can re-express inequality \eqref{Cabelloinequality} as
\beq
\alpha' \le 8.
\label{newinequality}
\eeq
Of course, rather than using Eq.~\eqref{jjj} to translate  \eqref{Cabelloinequality} from $\{-1,+1\}$-valued variables into $\{ 0,1\}$-valued variables, 
one can also just derive the inequality \eqref{newinequality} directly: among the $2^{18}$ possible assignments of values in $\{0,1\}$ to each of the $v_{\kappa}$, the maximum value of $\alpha'$ is 8.  Two examples of such assignments are provided in Fig.~\ref{LogicDefyingAssignments}.

It is useful to use a notation that specfies whether a given expectation value of some variable $X$ is relative to a preparation procedure $P$, in which case it is denoted $\langle X \rangle_P$,  or relative to an ontic state $\lambda$, in which case it is denoted $\langle X \rangle_{\lambda}$.  
We denote by $\alpha'(P)$ the quantity defined in \eqref{alphaprime} if the expectation values contained therein are relative to preparation $P$, and we denote by $\alpha'(\lambda)$ the case where the expectation values are relative to ontic state $\lambda$.
Under the assumption of an ontological model, each expectation value relative to a preparation $P$ can be expressed as a function of the expectation value relative to an ontic state $\lambda$, via
\beq
\langle X \rangle_P = \sum_{\lambda} \langle X \rangle_{\lambda} \;\mu(\lambda|P),
\label{previous}
\eeq
where $\mu(\lambda|P)$ is the distribution over ontic states associated with preparation $P$. 
We can infer from Eq.~\eqref{previous} that
\beq
\alpha'(P) = \sum_{\lambda} \alpha'(\lambda) \mu(\lambda|P).
\label{nnn}
\eeq

With these notational conventions, we can summarize the argument of Ref.~\cite{Cabelloexpt} as follows.  In any noncontextual ontological model, every ontic state $\lambda$ satisfies
\beq
\alpha'(\lambda) \le 8.
\label{lll}
\eeq
But this in turn implies, through Eq.~\eqref{nnn}, that for all preparations $P$,
\beq
\alpha'(P) \le 8,
\label{ooo}
\eeq
which is an inequality constraining operational quantities. 

We are now in a position to describe the problem with the inequality \eqref{ooo}, or equivalently inequality \eqref{Cabelloinequality}, and thus with the claim of Ref.~\cite{Cabelloexpt}.
First, we highlight the physical interpretation of the variables $v_{\kappa}$.  If $v_{\kappa}$ is assigned value 1 by the ontic state $\lambda$, then this means that if the system is in the ontic state $\lambda$, 
and  a measurement that includes $\kappa$ as an outcome is implemented on it, then the outcome $\kappa$ is certain to occur, while if $v_{\kappa}$ is assigned value 0 by $\lambda$, then the outcome $\kappa$ is certain {\em not} to occur.
But each of the $2^{18}$ different assignments to $(v_1 , \dots, v_{18})$  is such that for at least one measurement either:
{\em none} of the outcomes occur, as in the example of Fig.~\ref{LogicDefyingAssignments}(a), or {\em more than one} outcome occurs, as in the example of Fig.~\ref{LogicDefyingAssignments}(b).  (This is precisely what is implied by the fact that the 18 measurement events are {\em uncolourable}, as explained in 
the main text.)  Such assignments involve a {\em logical contradiction} given that the four outcomes of each measurement are mutually excusive and jointly exhaustive possibilities.

It follows that the sort of model that  a violation of inequality \eqref{ooo} 
rules out 
can already be ruled out {\em by logic alone}; no experiment is required. 
To put it another way, discovering that quantum theory and nature violate inequality \eqref{ooo}
only allows one to conclude that neither quantum theory nor nature involve a logical contradiction, which one presumably already knew prior to noting the violation.  

We have argued in the main text that the notion of KS-noncontextuality, insofar as it assumes outcome-determinism, is not suitable for devising experimentally robust inequalities given that every real measurement involves some noise.
The problem with inequality \eqref{ooo} can also be traced back to the use of the assumption of KS-noncontextuality.  Suppose we ask the following question: given the existence of nine four-outcome measurements satisfying the operational 
equivalences of Fig.~\ref{mmtequivs}(a), how are the operational probabilities that are assigned to these measurement events constrained if we presume that KS-noncontextual assignments underlie the operational statistics?
On the face of it, the question seems well-posed.  On further reflection, however, one sees that it is not.  There are simply {\em no} KS-noncontextual assignments to these measurement events, so it is simply impossible to imagine that such assignments could underlie the operational statistics.  There is nothing to be tested experimentally, as the hypothesis under consideration is seen to be false as a matter of logic.  

Here is another way to see that the inequality \eqref{ooo} does not provide a test of noncontextuality.  Consider the expectation value $\langle v_{\kappa_1} \oplus v_{\kappa_2} \oplus v_{\kappa_3} \oplus v_{\kappa_4} \rangle_{P} $ for a preparation $P$, where $\kappa_1$, $\kappa_2$, $\kappa_3$ and $\kappa_4$ correspond to the four outcomes of some measurement.   Regardless of which of the four outcomes of the measurement occurs in a given run where preparation $P$ is implemented---i.e. regardless of whether $(v_{\kappa_1} ,v_{\kappa_2} ,v_{\kappa_3}, v_{\kappa_4})$ comes out as (1,0,0,0) or (0,1,0,0) or (0,0,1,0) or (0,0,0,1) in that run---the variable $v_{\kappa_1} \oplus v_{\kappa_2} \oplus v_{\kappa_3} \oplus v_{\kappa_4} $ has the value 1.  We can think of it this way: the variable $v_{\kappa_1} \oplus v_{\kappa_2} \oplus v_{\kappa_3} \oplus v_{\kappa_4} $ is a trivial variable because it is a constant function of the measurement outcome. (This is analogous to how, in quantum theory, for a four-outcome 
measurement associated with four projectors, although each projector is a nontrivial observable, their sum is the identity operator, which has expectation value 1 for all quantum states, and therefore corresponds to a trivial observable.)  It follows that regardless of what distribution over the four outcomes is assigned by $P$, the expectation value $\langle v_{\kappa_1} \oplus v_{\kappa_2} \oplus v_{\kappa_3} \oplus v_{\kappa_4} \rangle_P$  will be 1.  Given that each of the nine terms in $\alpha'(P)$ is of this form, it follows that $\alpha'(P)=9$.  

So, for {\em any} operational theory that admits of nine four-outcome measurements with the operational equivalence relations depicted in Fig.~\ref{mmtequivs}(a), we will find that  $\alpha'(P)=9$ for all $P$.
Therefore,  we can conclude that the inequality $\alpha'(P)\le 8$ is violated for all $P$.  One can reach this conclusion without ever considering the question of whether the operational predictions can be explained by some underlying noncontextual model.  
  
Another consequence  of the triviality of the variables of the form $v_{\kappa_1} \oplus v_{\kappa_2} \oplus v_{\kappa_3} \oplus v_{\kappa_4} $ is that the inequality \eqref{ooo} can be violated regardless of how noisy the measurements are.  Suppose, for instance, that quantum theory describes our experiment, but that the nine four-outcome measurements are not the projective measurements described in Fig.~\eqref{CEGAhypergraph}, but rather noisy versions thereof.
For instance, one can imagine that each measurement 
is associated with a positive operator-valued measure that is the image under a depolarizing map of the projector valued measure associated with the ideal measurement.
The amount of depolarization can be taken arbitrarily large and, as long as it is the {\em same} amount of depolarization for each of the measurements, the nine noisy measurements that result will still satisfy precisely the same operational equivalences as the original nine, namely, those depicted in Fig.~\ref{mmtequivs}(a).   For such noisy measurements, we can still identify variables $v_{\kappa}$ associated to the eighteen equivalence classes of measurement events, and we still find that regardless of which of the four outcomes of the measurement occurs, the variable $v_{\kappa_1} \oplus v_{\kappa_2} \oplus v_{\kappa_3} \oplus v_{\kappa_4} $ has the value 1, so that
regardless of what distribution over the four outcomes is assigned by $P$, the expectation value $\langle v_{\kappa_1} \oplus v_{\kappa_2} \oplus v_{\kappa_3} \oplus v_{\kappa_4} \rangle_P$  will be 1 and therefore $\alpha'(P)=9$, which is a violation of the inequality \eqref{ooo}.

According to the generalized notion of noncontextuality proposed in Ref.~\cite{Spe05}, if one adds enough noise to the 
preparations and measurements in an experiment, it always becomes possible to represent the experimental statistics by a
noncontextual model.  One way to prove this is to note that: (i)  if all of the preparations and the measurements in an 
experiment admit of positive Wigner representations, then, as demonstrated in Ref.~\cite{Spe08}, the Wigner representation
defines a noncontextual model, and (ii) if one adds enough noise to the preparations and measurements, it is possible to 
ensure that they admit of positive Wigner representations.

This analysis of the effect of noise accords with intuition: noncontextuality is meant to represent a notion of classicality, so that a failure of noncontextuality is only expected to occur in a quantum experiment if one's experimental operations have a high degree of coherence.
It follows that  there should always exist a threshold of noise above which an experiment cannot be used to demonstrate the failure of noncontextuality.   
One can turn this observation into a minimal criterion that should be satisfied by any noncontextuality inequality: there should exist a threshold of experimental noise above which a noncontextuality inequality cannot be violated. 

As we have just noted, the inequality proposed in Ref.~\cite{Cabelloexpt} fails this minimal criterion. By contrast, the noncontextuality inequality proposed in this article identifies such a threshold for the 18 ray construction:
the noise must be kept low enough that the average of the measurement predictabilities is above $5/6$.

\subsection{Alternative interpretation}

The inequality proposed in Ref.~\cite{Cabelloexpt} 
can be given a different interpretation to the one provided in the previous subsection.  
This  interpretation is more charitable in some ways, but it still does not vindicate the proposed inequality as delimiting the boundary of noncontextual models. 

\begin{figure}[htb]
 \includegraphics[scale=0.3]{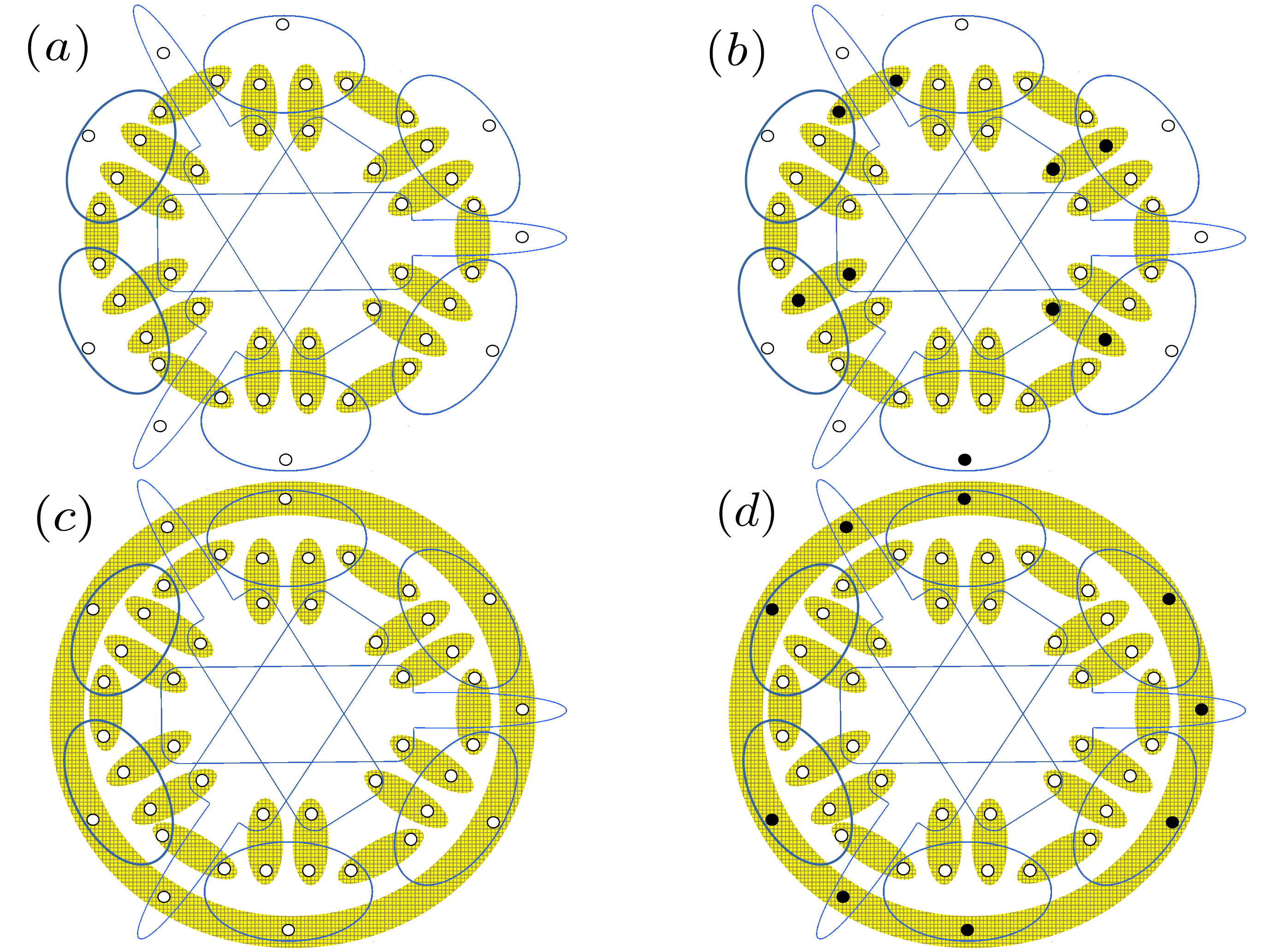}
 \caption{(a) The hypergraph wherein each measurement is assigned an additional fifth outcome.  (b) A normalized noncontextual deterministic assignment to the hypergraph of (a) that recovers the subnormalized noncontextual deterministic assignment of Fig.~\ref{LogicDefyingAssignments}(a) on the appropriate subgraph; (c) The hypergraph wherein the fifth outcomes are all operationally equivalent; (d) the unique normalized noncontextual and deterministic assignment to the hypergraph of (c).}
\label{Subnormalization}
\end{figure}

The idea is to imagine that for each of the nine measurements, there are in fact {\em five} rather than four outcomes that are mutually exclusive and jointly exhaustive. 
Thus, in this interpretation, it is assumed that the hypergraph describing compatibility relations and operational equivalences is {\em not} the one of Fig.~\ref{mmtequivs}(a), but rather a modification wherein there are nine additional nodes---one additional node appended to each of the nine measurements---as depicted in Fig.~\ref{Subnormalization}(a).

If $\{ \kappa_1, \kappa_2, \kappa_3, \kappa_4\}$  are the original four outcomes of a given measurement, then the variable $v_{\kappa_1} \oplus v_{\kappa_2} \oplus v_{\kappa_3} \oplus v_{\kappa_4} $ is no longer a constant function of the measurement outcome because its value varies depending on whether or not the fifth outcome occurs.  If $\kappa_5$ denotes the fifth outcome of the measurement, then the trivial variable is $v_{\kappa_1} \oplus v_{\kappa_2} \oplus v_{\kappa_3} \oplus v_{\kappa_4} \oplus v_{\kappa_5}$, taking the value 1 regardless of the outcome.

In this case, the assignments of the type depicted in Fig.~\ref{LogicDefyingAssignments}(a)---the noncontextual deterministic assignments that are {\em subnormalized}---can be embedded into noncontextual deterministic {\em normalized} assignments on the larger hypergraph, as depicted in Fig.~\ref{Subnormalization}(b).  (The possibility of such an embedding for the subnormalized noncontextual deterministic assignments considered in Cabello, Severini and Winter~\cite{CSW} was noted in Acin, Fritz, Leverrier, Sainz~\cite{AFLS}.)

Of course, such a move does not provide any way of understanding the deterministic noncontextual assignments of the type depicted in Fig.~\ref{LogicDefyingAssignments}(b), because the latter violate normalization by having the probabilities of the different outcomes of the measurement summing to greater than 1---they are {\em supernormalized}.  

So, while the supernormalized noncontextual deterministic assignments can be ruled out by logic alone, the subnormalized noncontextual deterministic assignments may be entertained without logical inconsistency if they are considered as reductions to a subgraph of a normalized noncontextual deterministic assignment on a larger hypergraph.

Because the justification given in Ref.~\cite{Cabelloexpt} for the inequality derived there asks one to consider {\em all} of the noncontextual deterministic assignments, including the supernormalized ones, the interpretation of this inequality as a constraint on subnormalized assignments is in tension with the manner in which the inequality is justified.
This interpretation is a better fit with Cabello's later work, such as Ref.~\cite{CSW}, wherein the restriction to subnormalized assignments is explicit. 
In any case,  if the inequality holds for {\em all} noncontextual deterministic assignments, regardless of normalization, then it holds for the special case of the subnormalized assignments, so the inequality can still be derived within this interpretation.

The problem with this interpretation 
becomes manifest when we require that  the original hypergraph of Fig.~\ref{mmtequivs}(a)---and thus the corresponding subgraph of Fig.~\ref{Subnormalization}(a) from which it is derived in this interpretation---is realized in terms of Hilbert-space bases in the manner depicted in Fig.~\ref{CEGAhypergraph}(a).

We consider two possible ways of fulfiling this requirement, and explain why it is not possible to vindicate the inequality of Eq.~\eqref{newinequality} in either case.

In one approach, we imagine that the quantum
system is in fact described by a 5-dimensional Hilbert space.  In this case, rank-1 projective measurements have five outcomes and are therefore described within the hypergraph representation by an edge with five 
nodes, just as we have for the measurements in Fig.~\ref{Subnormalization}(a).
Now consider an association of Hilbert space rays with the nodes of this hypergraph such that one recovers the association of rays to nodes described by Fig.~\ref{CEGAhypergraph} on the subgraph of Fig.~\ref{Subnormalization}(a) that corresponds to the original hypergraph of Fig.~\ref{mmtequivs}(a).  This is possible if, for every measurement, the fifth outcome is associated with a ray that is orthogonal to the 4d subspace in which all of the other rays live.  
But then, under a tomographically complete set of preparations of the 5d Hilbert space, one finds that the fifth outcomes are all operationally equivalent, so that the appropriate hypergraph is not that of Fig.~\ref{Subnormalization}(a) but rather the one depicted in Fig.~\ref{Subnormalization}(c).  

Now, consider {\em this} hypergraph.  It only admits of a single normalized noncontextual deterministic assignment, the one that assigns 0s to every outcome in the original set  and 1 to all of the fifth outcomes, as depicted in Fig.~\ref{Subnormalization}(d).  Therefore, if one were to experimentally verify the applicability of the hypergraph of Fig.~\ref{Subnormalization}(c), by verifying the operational equivalences depicted therein, then any KS-noncontextual model consistent with this hypergraph would not only satisfy the inequality $\alpha'(\lambda)\le 8$ (Eq.~\eqref{lll}), it would predict that {\em all} of the measurement events appearing in the inequality receive value 0, so that the inequality could be 
strengthened to the equality $\alpha'(\lambda)=0$, which in turn would imply, through Eq.~\eqref{nnn}, that for all preparation procedures $P$, the operational inequality $\alpha'(P)\le 8$ could be strengthened to the operational equality
\beq
\alpha'(P)=0.
\eeq  
But this is trivial to violate experimentally: simply find a preparation that does not always yield the fifth outcome for every measurement.   

We take the triviality of this constraint to speak against the idea that it captures the assumption of noncontextuality. 
Therefore, the conclusion to draw from this discussion is {\em not} that one should replace the inequality $\alpha'(P)\le 8$  with $\alpha'(P)=0$.
Rather, as we've argued at length in the main text, because the KS-noncontextual models make the unjustified assumption of outcome-determinism, the notion of noncontextuality should not be formalized as KS-noncontextuality, but rather as measurement and preparation noncontextuality. 

We now turn to the second approach.  Here, one sticks to the notion that the quantum system being probed is 4-dimensional and instead one suggests
that each of the nine measurements is nonprojective, that is, each is represented by a positive operator valued measure rather than a projector valued measure.  In this way, one can ensure that the measurements indeed have five outcomes. 
One might even think of the fifth outcome as representing a `no detection' event (the idea of justifying subnormalized assignments by imagining an additional `no detection' outcome has also been discussed in Ref.~\cite{AFLS}).

To see that there is something fishy about this approach, it suffices to note that if it were correct, then it would have the bizarre consequence that
in the case where the measurements achieve the ideal of projectiveness, satisfaction of the  inequality $\alpha'(P) \le 8$ is ruled out by logic alone,
whereas if the measurements depart from this ideal, however little, suddenly the inequality specifies whether or not the experiment can be modelled noncontextually.

In any case, the real problem with this approach is easily identified. 
For a nonprojective measurement, one is assigning probabilities to {\em effects} (positive operators less than identity) rather than projectors.  In this case, one must allow noncontextual assignments to be {\em probabilistic}.  This has been proven elsewhere~\cite{Spe14} and we will not repeat the arguments here.  Such probabilistic noncontextual assignments are not restricted to be in the convex hull of the deterministic noncontextual assignments, and therefore can be more general than mixtures of the latter.  
Because the derivation of the inequality $\alpha'(P) \le 8$ made crucial use of the assumption that the preparation $P$ was a mixture of deterministic noncontextual assignments, the fact that the assumption of determinism is unwarranted implies that one can no longer derive the inequality as a constraint on noncontextual models.

\begin{thebibliography}{5}
\bibitem{belldeterminism}  Indeed, in Ref.~\cite{Bellbook} (p. 157), Bell writes
\begin{quote}
My own first paper on [the subject of Bell's Theorem] ... starts with a summary of the EPR argument from locality to deterministic hidden variables. But the commentators have almost universally reported that it begins with deterministic hidden variables.
\end{quote}
Although Wiseman has disputed Bell's account of the role of determinism in his first paper\cite{Wiseman}, see Norsen's response\cite{Norsen}.
\bibitem{merminquote} Indeed, in Ref.~\cite{Cabello98}, David Mermin is quoted as having said: 
``the whole notion of an experimental test of [B]KS misses the point'', a view that was held by many researchers at the time.
\end{thebibliography}
\end{document}